\documentclass[journal=cmatex,manuscript=article]{achemso}

\usepackage{natbib}
\usepackage{adjustbox}
\usepackage{xcolor, hyperref}
\usepackage[version=3]{mhchem} % Formula subscripts using \ce{}

\author{Rohit Batra}
\affiliation{Department of Materials Science \& Engineering, University of Connecticut, Storrs, CT, USA}

\author{Tran Doan Huan}
\affiliation{Department of Materials Science \& Engineering, University of Connecticut, Storrs, CT, USA}

\author{George A. Rossetti, Jr.}
\affiliation{Department of Materials Science \& Engineering, University of Connecticut, Storrs, CT, USA}

\author{Rampi Ramprasad}
\email{rampi.ramprasad@uconn.edu}
\affiliation{Department of Materials Science \& Engineering, University of Connecticut, Storrs, CT, USA}
\title{Dopants Promoting Ferroelectricity in Hafnia: Insights From A Comprehensive Chemical Space Exploration}

\begin{document}
%%%%%%%%%%%%%%%%%%%%%%%%%%%%%%%%%%%%%%%%%%%%%%%%%%%%%%%%%%%%%%%%%%%%%
%% The manuscript does not need to include \maketitle, which is
%% executed automatically.  The document should begin with an
%% abstract, if appropriate.  If one is given and should not be, the
%% contents will be gobbled.
%%%%%%%%%%%%%%%%%%%%%%%%%%%%%%%%%%%%%%%%%%%%%%%%%%%%%%%%%%%%%%%%%%%%%
\newpage
\begin{abstract}
Although dopants have been extensively employed to promote ferroelectricity in hafnia films, their role in stabilizing the responsible ferroelectric non-equilibrium $Pca$2$_1$ phase is not well understood. In this work, using first principles computations, we investigate the influence of nearly 40 dopants on the phase stability in bulk hafnia to identify dopants that can favor formation of the polar $Pca$2$_1$ phase. Although no dopant was found to stabilize this polar phase as the ground state, suggesting that dopants $\it{alone}$ cannot induce ferroelectricity in hafnia, Ca, Sr, Ba, La, Y and Gd were found to significantly lower the energy of the polar phase with respect to the equilibrium monoclinic phase. These results are consistent with the empirical measurements of large remnant polarization in hafnia films doped with these elements. Additionally, clear chemical trends of dopants with larger ionic radii and lower electronegativity favoring the polar $Pca$2$_1$ phase in hafnia were identified. For this polar phase, an additional bond between the dopant cation and the 2nd nearest oxygen neighbor was identified as the root-cause of these trends. Further, trivalent dopants (Y, La, and Gd) were revealed to stabilize the polar $Pca$2$_1$ phase at lower strains when compared to divalent dopants (Sr and Ba). Based on these insights, we predict that the lanthanide series metals, the lower half of alkaline earth metals (Ca, Sr and Ba) and Y as the most suitable dopants to promote ferroelectricity in hafnia.
\end{abstract}

\newpage
\section{Introduction}
Intentionally added impurities, i.e., dopants, can completely alter the physical properties of the host material. While in some cases, the additional electrons or holes contributed by the dopants dramatically modify the electronic structure, thereby changing properties like the electrical conductivity\cite{doped_semiconductor} and magnetism\cite{DMS}, in other cases, the small doping-induced perturbation is enough to alter the atomic arrangement (crystal structure) of the host system (e.g. yttrium stabilized zirconia). Hafnia (HfO$_2$), a well known linear dielectric material\cite{hafnia_review_Hong_Zhu, hafnia_high_k_JAP_review, hafnia_high_k_progress_physics_Robertson, hafnia_high_k_rampi, hafnia_high_k_tang}, is likely an example of the latter, as doped thin films of this material have been recently observed to exhibit ferroelectric (FE) behavior through the formation of a non-equilibrium polar phase.\cite{hafnia_ferro_observation, undoped_hafnia} Despite a great number of experimental and theoretical studies,\cite{Huan_ferroelectricity, hafnia_surface_energy_Materlik,hafnia_surface_energy_Rohit,hafnia_mono_to_tetra_surface_energy} the origin of this novel functionality, which has applications in FE-field effect transistors\cite{device_fet} and FE-random access memories,\cite{device_feram} has not been completely understood.

In the most likely mechanism, some ``suitable'' combination of surface energy, mechanical stresses, oxygen vacancies, dopants and the electrical history of the hafnia film is believed to stabilize the polar orthorhombic Pca2$_1$ (P-O1) phase over the equilibrium monoclinic (M) phase of hafnia, thus enabling FE behavior.\cite{hafnia_review, hafnia_dopants_effects, dopants_hafnia_influence_RSC, TEM_PO1_observation_hafnia} The disappearance of ferroelectricity in the absence of a capping electrode and with increasing film thickness suggests the critical role of the mechanical stresses\cite{hafnia_ferro_observation, stress_hafnia_rohit, hzo_strain_Min_Park, zirconia_stress_Kisi, mechanical_stress_Al, mechanical_stress_Y} and surface energies,\cite{hafnia_surface_energy_Materlik, hafnia_surface_energy_Rohit, hafnia_mono_to_tetra_surface_energy}respectively. Similarly, the demonstration of the ``wake-up effect'' (on application of external electric fields) hints at the role that the electrical history of the film plays in stabilizing the FE phase.\cite{hystereses_deform_Tony, electric_field_ovac_movement, electric_field_structure} Dopants, too, have been found to increase the stability ``window'' of the P-O1 phase as reflected in an increase in both the magnitude of the measured polarization and the critical thickness of the hafnia film (below which FE behavior is observed).\cite{undoped_hafnia} Some insight into the role of dopants has emerged from recent empirical studies,\cite{dopants_hafnia_influence_RSC,hafnia_dopants_effects} which have indicated the trend of dopants with higher ionic radii leading to enhanced polarization. Nevertheless, the true role of the dopants in the formation of the P-O1 phase remains unclear, given that traditionally doping is known to stabilize the high-temperature tetragonal (T) or the cubic phases of hafnia.\cite{dopants_stabilize_T_or_C, dopants_stabilize_T, hafnia_hong_design_dopants} Two critical questions, important from both application and theoretical standpoints, that these recent studies\cite{hafnia_ferro_observation, mechanical_stress_Al, mechanical_stress_Y, dopants_hafnia_influence_RSC,hafnia_dopants_muller} on FE doped hafnia raise are: (1) which dopant favor the polar phase the most and at what concentration?, and (2) do dopants play a critical role in stabilizing this polar phase in hafnia films, and if yes, which attributes of a dopant (chemical or physical) are relevant?

\begin{figure}
\centering
\includegraphics[scale=0.57]{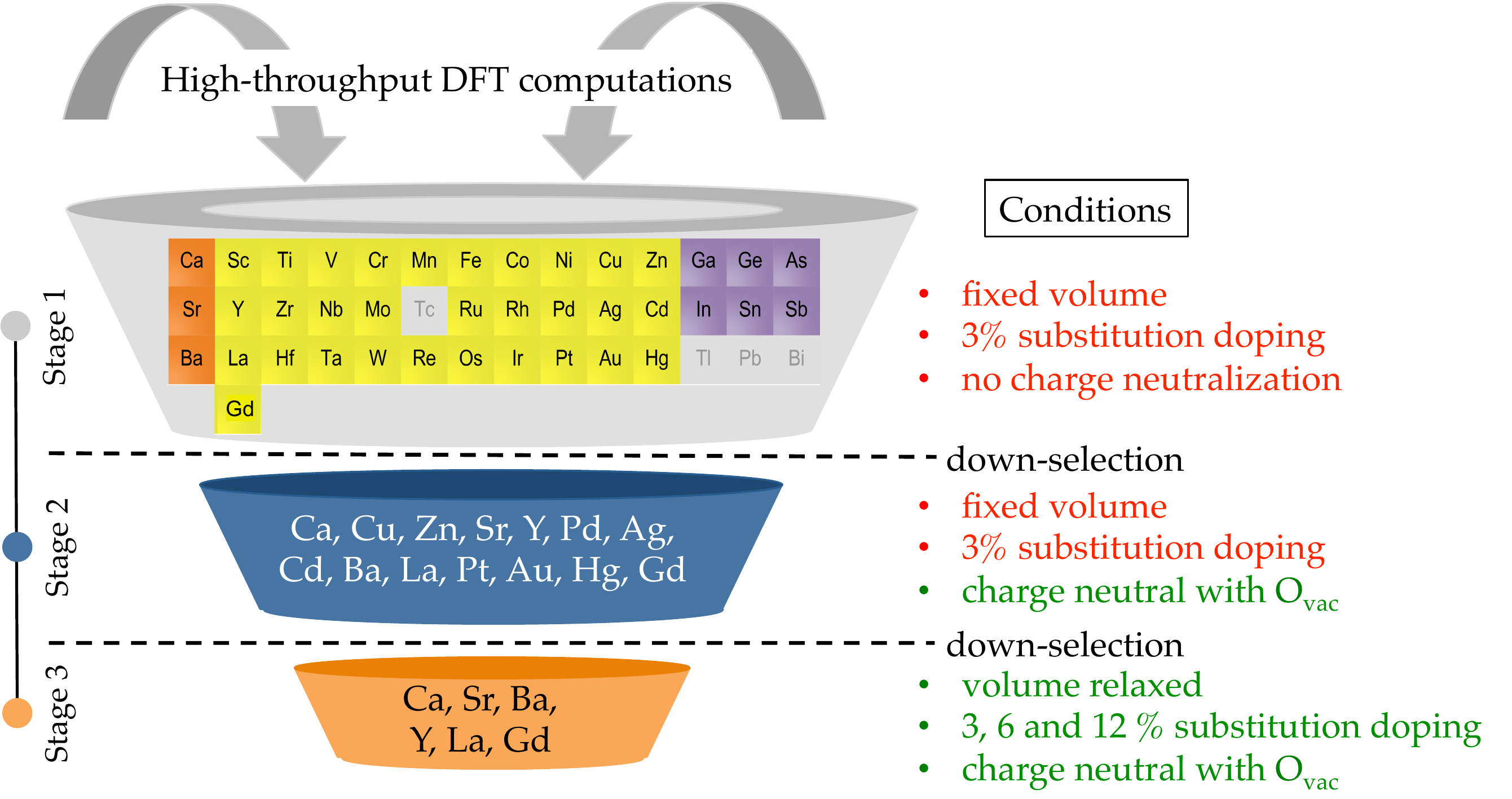}
\caption{The overall scheme of this work illustrating the three-stage selection process and the modeling conditions imposed in each stage.}
\label{Fig:scheme}
\end{figure}

In this contribution, we address these questions using high-throughput first-principles density functional theory (DFT) computations. In order to address the first question, we follow a three-stage down-selection strategy, illustrated in Fig. \ref{Fig:scheme}, wherein we examine the influence of nearly 40 dopants on the energetics of the relevant low-energy phases of hafnia, including M ($P2_1/c$), T ($P4_2/nmc$), P-O1 ($Pca2_1$), another polar P-O2 ($Pmn2_1$), and high-pressure OA ($Pbca$) phases. Based on these energy changes, the initial set of nearly 40 dopants in Stage 1 is down-selected to 14 dopants in Stage 2, and finally, to the 6 most promising dopants, i.e., Ca, Sr, Ba, Y, La and Gd, in Stage 3. In agreement with empirical observations,\cite{hafnia_dopants_effects, dopants_hafnia_influence_RSC} our study revealed that these 6 dopants favor the stabilization of the P-O1 phase of hafnia. To answer the second question, the computational data obtained in Stage 3 was analyzed. Clear trends illustrating that dopants with higher ionic radii and lower electronegativity stabilize the P-O1 phase the most were found, also consistent with the experimental observations.\cite{dopants_hafnia_influence_RSC} The root-cause of these trends is traced to the formation of an additional bond between the dopant and the 2$^{\rm nd}$ nearest-neighbor oxygen atom. Based on these findings, we search the entire Periodic Table, predicting the lanthanides, the lower half of the alkaline earth metals (i.e. Ca, Sr, Ba) and Y as the most favorable dopants to promote ferroelectricity in hafnia.

\section{Theoretical Methods}
Our work is based on electronic structure DFT calculations, performed using the Vienna {\it Ab Initio} Simulation Package\cite{vasp} (VASP) employing the Perdew-Burke-Ernzerhof exchange-correlation functional\cite{PBE} and the projector-augmented wave methodology.\cite{PAW} A 3$\times$3$\times$3 Monkhorst-Pack mesh\cite{monkhorst} for k-point sampling was adopted and a basis set of plane waves with kinetic energies up to 500 eV was used to represent the wave functions. For each doped phase, spin polarized computations were performed and all atoms were allowed to relax until atomic forces were smaller than 10$^{-2}$ eV/\AA.

To determine the energy ordering of phases in doped hafnia, we define the relative energy of a phase $\alpha$ with respect to the equilibrium M phase in the presence of a dopant D as
\begin{equation}\label{eq:rel_energy}
\Delta E^{\rm \alpha-M}_{\rm D} = E^\alpha_{\rm D} - E^{\rm M}_{\rm D},
\end{equation}
where $E^\alpha_{\rm D}$ and $E^{\rm M}_{\rm D}$ are the DFT computed energies of the doped $\alpha$ and M phases, respectively. To highlight the direct role of a dopant in stabilizing the phase $\alpha$, we subtract from Eq. \ref{eq:rel_energy} a term corresponding to the energy of dopant-free pure phases:
\begin{equation}\label{eq:rel_rel_energy}
\Delta E^{\alpha-{\rm M}}_{\rm D-Pure} = \left(E^\alpha_{\rm D} - E^{\rm M}_{\rm D}\right) -\left(E^\alpha_{\rm Pure} - E^{\rm M}_{\rm Pure}\right)
\end{equation}
where $E^\alpha_{\rm Pure}$ and $E^{\rm M}_{\rm Pure}$ are the DFT computed energies of pure $\alpha$ and M phases, respectively. $\Delta E^{\rm \alpha-M}_{\rm D-Pure}$ represents the change in the relative energy of the phase $\alpha$ with respect to the M phase solely due to the introduction of the dopant D. Thus, a dopant with negative $\Delta E^{\rm \alpha-M}_{\rm D-Pure}$ favors (or stabilizes) the phase $\alpha$ over the M phase more than in the dopant-free pure case. Further, if $\alpha$ happens to be one of the polar phases, one can expect such dopants to enhance FE behavior in hafnia.

Five different phases of hafnia were considered, including M, T, P-O1, P-O2 and OA, as they were either empirically observed or theoretically predicted to have low energy under conditions for which hafnia films display FE behavior.\cite{Huan_ferroelectricity,Hafnia_ph_dig} Equivalent 32 formula-unit (96 atom) supercells, starting from the structures documented in our previous work,\cite{stress_hafnia_rohit} were constructed to carry out the energy calculations. For each phase, three levels of substitutional doping concentration, namely, 3.125\%, 6.25\% and 12.5\% were studied by replacing 1, 2 and 4 Hf atom(s), respectively, by the dopant atom(s).

To overcome the challenge of high computational cost associated with accurately modeling the effect of $\sim$40 dopants on the energetics of the five phases of hafnia, we carry out this work in three stages, as illustrated in Fig. \ref{Fig:scheme}. Moving down the stages, a balance between computational accuracy and cost is maintained by increasing the modeling sophistication on the one hand and retaining only the promising dopants, with substantially negative $\Delta E^{\rm PO1-M}_{\rm D-Pure}$ and $\Delta E^{\rm PO2-M}_{\rm D-Pure}$, on the other hand. We restrict the initial set of dopants to elements from row 3, 4 and 5 of the Periodic Table (see Fig. \ref{Fig:scheme}), with the exception of Gd, which is included since empirical observations of ferroelectricity have been made in this case. In Stage 1, we model these dopants in the aforementioned five phases at 3.125\% doping concentration, and under the assumption of fixed volume of the simulation cell and the absence of oxygen vacancies (O$_{\rm vac}$). The relatively large size of the dopants considered and small perturbations expected at such small doping concentration form the rationale underlying these assumptions. Promising dopants from Stage 1 that energetically favor the polar phases were selected for more in-depth studies in Stage 2. Their influence on the phase stability was again studied at the doping concentration of 3.125\%, but now in a presence of appropriate concentration of O$_{\rm vac}$ (determined through the study of the electronic structures of doped hafnia phases, as discussed in Supplementary Information), expected to be present in real systems owing to the different oxidation states of the dopant and the hafnium ion. Finally, in Stage 3, promising dopants selected from Stage 2 were studied at multiple doping concentrations of 3.125\%, 6.25\% and 12.5\%. The volume of the supercell was relaxed and an appropriate number of O$_{\rm vac}$ were introduced to achieve charge neutrality. The doped hafnia structures obtained in Stage 3 were later examined to draw key chemical trends.

\section{Results and Discussion}
\subsection{Stage 1}
\begin{figure}[h]
	\centering
	\includegraphics[scale=0.725]{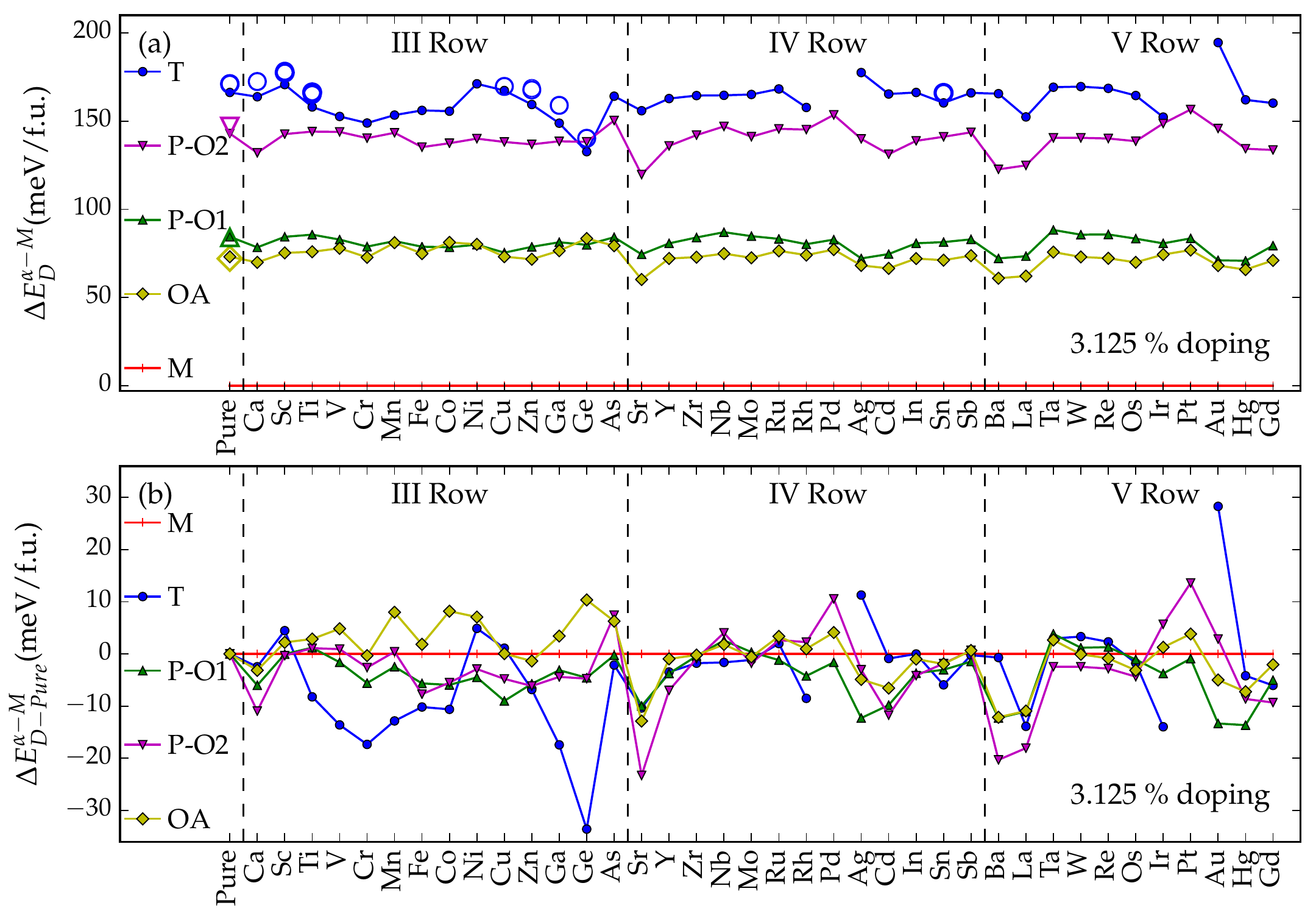}
	\caption{Phase stability of hafnia in presence of different dopants and under the constraints of Stage 1, as computed using (a) Eq. \ref{eq:rel_energy} and (b) Eq. \ref{eq:rel_rel_energy}. In panel (a), solid symbols represent the data from this work while open symbols signify results from previous studies.\cite{dopants_stabilize_T_or_C,Huan_ferroelectricity} The lines are guide to the eyes.}
	\label{Fig:stage1}
\end{figure}
As stated above and illustrated in Fig. \ref{Fig:scheme}, the influence of $\sim$40 dopants on the phase stability in hafnia under the assumption of fixed volume and the absence of O$_{\rm vac}$ was studied in Stage 1. The energies of different phases of hafnia at 3.125\% doping concentration are presented in Fig. \ref{Fig:stage1} and are found to be consistent with limited available past studies (shown in open circles).\cite{Huan_ferroelectricity,dopants_stabilize_T_or_C} In case of pure hafnia, the small energy difference between the equilibrium M and the P-O1 phases should be noted, signaling that even minor perturbations, perhaps introduced by extrinsic factors, such as dopants, stresses, etc., may be sufficient to stabilize the polar P-O1 phase as the ground state. Further, the P-O1 and the OA phases are extremely close in energy, in agreement with the previous studies.\cite{hafnia_LDA_vs_GGA, Huan_ferroelectricity, hafnia_surface_energy_Materlik} This energetic proximity is a manifestation of the remarkable structural similarity between the two phases.

As captured in Fig. \ref{Fig:stage1}(a), the M phase remains the equilibrium phase for all the dopants considered at 3.125\% doping concentration, although the energy differences among the hafnia phases change significantly. The relative energy of T phase alters substantially more with the choice of the dopant (for e.g., Ge, Au, etc.) in comparison to that of the P-O1, P-O2 and OA phases, possibly due to the different coordination environment experienced by a dopant cation in the T (CN = 8) versus the other phases (CN = 7) considered here. Interestingly, the T phase of Pd- and Pt-doped hafnia collapse into the P-O1 phase (see Supplementary Information for details) upon atomic relaxation (resulting in absence of these data points in Fig. \ref{Fig:stage1}). An important implication of this finding is that even small perturbations can possibly result in T to P-O1 phase transformations, and can be a potential pathway of formation of the P-O1 phase in hafnia. We will continue to encounter this collapse of the T phase to the P-O1 phase in later stages of this work as well.

Owing to the large energy scale and the small doping level, the influence of dopants on the phase stability appears feeble in Fig. \ref{Fig:stage1}(a). This picture, however, changes substantially when we re-plot it using Eq. \ref{eq:rel_rel_energy} as shown in Fig. \ref{Fig:stage1}(b). We again caution here that the quantity $\Delta E_{\rm D-Pure}^{\rm \alpha-M}$ plotted in Fig. \ref{Fig:stage1}(b) only helps us identify the phase(s) a dopant prefers over the M phase, and not the lowest energy ground state of hafnia, which is indeed determined by the quantity $\Delta E_{\rm D}^{\rm \alpha - M}$. Two key trends to be observed in Fig. \ref{Fig:stage1}(b) are: (1) row IV and row V dopants follow very similar phase stability trends when moving from left to right across the periodic table, with the row V dopants inducing larger energy variations, and (2) dopants from alkaline earth, and group 3, 10, 11 and 12 of the periodic table tend to favor the P-O1 and/or the P-O2 phases in hafnia, leading to the following shortlisted candidates further studied in Stage 2: Ca, Sr, Ba, Y, La, Cu, Zn, Pd, Ag, Cd, Pt, Au, Hg and Gd. Interestingly, a few of these dopants, such as Y, La, Sr, Ba, La, among others, have been empirically\cite{dopants_hafnia_influence_RSC,hafnia_dopants_effects} shown to promote substantial FE behavior in hafnia films, thus, already highlighting an agreement between our initial results and experiments. Another vital chemical insight, which will be strengthened in the later sections, is that dopants with low electronegativity tend to stabilize the polar phases in hafnia.

\subsection{Stage 2}
In Stage 2, we increase the modeling sophistication by introducing appropriate charge neutralizing O$_{\rm vac}$ for 3.125\% doped hafnia systems. Two issues concerning the number of O$_{\rm vac}$ and their placement site in the 32 hafnia-unit supercell should be addressed. Since all the dopants, except Y, La, Au and Gd, in Stage 2 are divalent, only one O$_{\rm vac}$ corresponding to the one dopant cation needs to be added (as confirmed using the electronic structure studies discussed in Supplementary Information). However, for the case of Y, La, Au and Gd, a partial O$_{\rm vac}$ is required at 3.125\% doping level. To avoid practical computational issues, these trivalent dopants were transferred directly to Stage 3. The remaining 10 divalent dopants were studied in Stage 2 with a single O$_{\rm vac}$.

\begin{figure}
	\centering
	\includegraphics[scale=0.725]{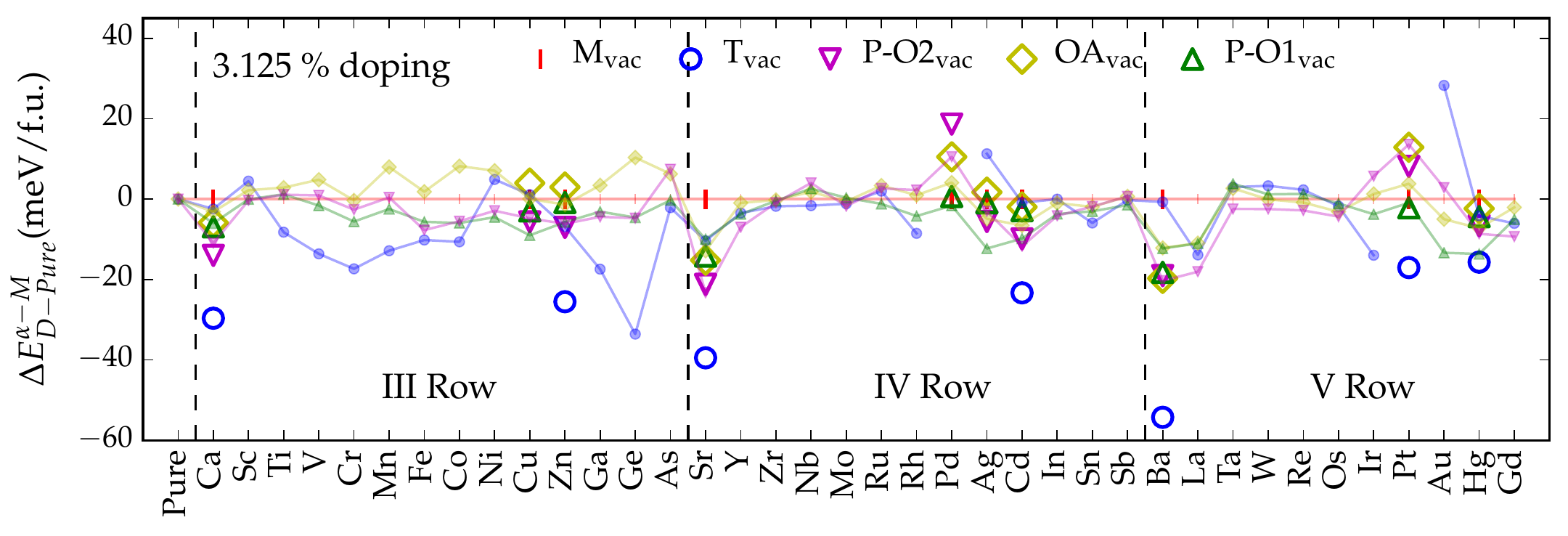}
	\caption{The relative energies of 3.125\% doped hafnia for the limited set of 10 divalent dopants of Stage 2 in presence of a charge neutralizing O$_{\rm vac}$. For ease of comparison, the results of Stage 2 (open symbols) are overlaid on top of that of Stage 1 (lighter solid symbols).}
	\label{Fig:stage2}
\end{figure}

With respect to the placement of this single O$_{\rm vac}$, we argue that this should be in a nearest-neighbor site to the dopant cation owing to the electrostatic pull expected between the negatively charged dopant and the positively charged O$_{\rm vac}$ defects. With this restriction on configurational space to the cases in which O$_{\rm vac}$ is closest to the dopant, and taking into account the symmetry of the different hafnia phases, we are left with 7 different choices for the M, P-O1, and OA phases, 5 for the O2 and 2 for the T phase. These choices can be further classified into two categories based on the number of Hf-O bonds that need to be broken to introduce an O$_{\rm vac}$; while one category involves breaking 3 bonds, the other requires 4 broken bonds. For the representative case of Pd- and Pt- doped hafnia systems, energies for all possible configurations (i.e., 7 for the M, P-O1, and OA, 5 for the O2 and 2 for the T) were computed and it was found that O$_{\rm vac}$ sites involving 3 broken Hf-O bonds are always energetically preferred, with the exception of the T phase which has only one type of O$_{\rm vac}$ site that involves breaking 4 Hf-O bonds. Thus, we further reduce our configurational space to cases which involve breakage of only 3 Hf-O bonds in the M, P-O1, OA and P-O2 phases. This leaves us with 3 different choices for the M, P-O1, OA phases, and 2 choices for each of the O2 and T phases. For each phase, only the configuration with lowest energy was considered in order to obtain the phase stability trends presented in Fig. \ref{Fig:stage2}. To summarize, in Stage 2 we computed the phase stability of hafnia at dopant concentration of 3.125\% for the case of the 10 shortlisted divalent elements, and with the restrictions of O$_{\rm vac}$ being in nearest-neighbor site of the dopant and occupying an O site with 3 Hf-O bonds in the case of M, P-O1, OA and O2 phases. The volume of the supercell was also assumed to be fixed.

The findings of Stage 2 are overlaid on the results of Stage 1 for the selected set of 10 divalent dopants in Fig. \ref{Fig:stage2}. The transition metals that favored the polar phase(s) in Stage 1, do not substantially stabilize the polar phase(s) with the introduction of O$_{\rm vac}$ as $\Delta E_{\rm D-Pure}^{\rm \alpha-M}$ of both the polar phases can be seen to shift up after the O$_{\rm vac}$ introduction (e.g., compare the open and solid symbols for the case of Cu and Zn in Fig. \ref{Fig:stage2}). On the other hand, the T phase is consistently favored with the addition of O$_{\rm vac}$ due to the lowering of the coordination number of the vacancy neighboring Hf atoms from 8 to 7, which is energetically preferred - and is also the reason why the M phase is the equilibrium phase of hafnia. This behavior is consistent with the past study\cite{hafnia_vacancy_stabilization}. The Cu- and Ag-doped T phase was, however, found to collapse into the polar P-O1 phase. Further investigations are necessary to identify what triggers this collapse of the T phase into the P-O1 phase. Nevertheless, the alkaline earth metals like Ca, Sr, and Ba continue to stabilize the polar phases in Stage 2, leaving us with our next set of promising candidates studied in Stage 3, i.e, Ca, Sr, Ba, Y, La, Au and Gd.

\subsection{Stage 3}
From the initial set of $\sim$40 dopants, we are now left with the 7 most promising candidates in Stage 3 that favor the polar phase(s) in hafnia. Owing to the lesser number of dopants involved, we now lift the modeling constraints imposed in the previous stages, and investigate the influence of these dopants at varying concentrations. For the case of divalent dopants, we studied three different doping concentrations of 3.125\%, 6.25\% and 12.5\%. On the other hand, for the case of trivalent dopants, we studied only 6.25\% and 12.5\% doping concentrations owing to the difficulty associated with modeling a partial O$_{\rm vac}$ at 3.125\% doping level, as mentioned earlier. The volume of the supercell was relaxed and an appropriate number of O$_{\rm vac}$ were introduced to achieve charge neutrality. Unfortunately, for the case of Au, the phases did not retain their structural identity (i.e., the relaxed structures from our computations were so distorted that they could not be unambiguously associated with the starting structure) at higher doping concentration of 6.25\% and 12.5\%, and thus, we exclude this case from our results.
\begin{figure}[h]
	\centering
	\includegraphics[scale=0.78]{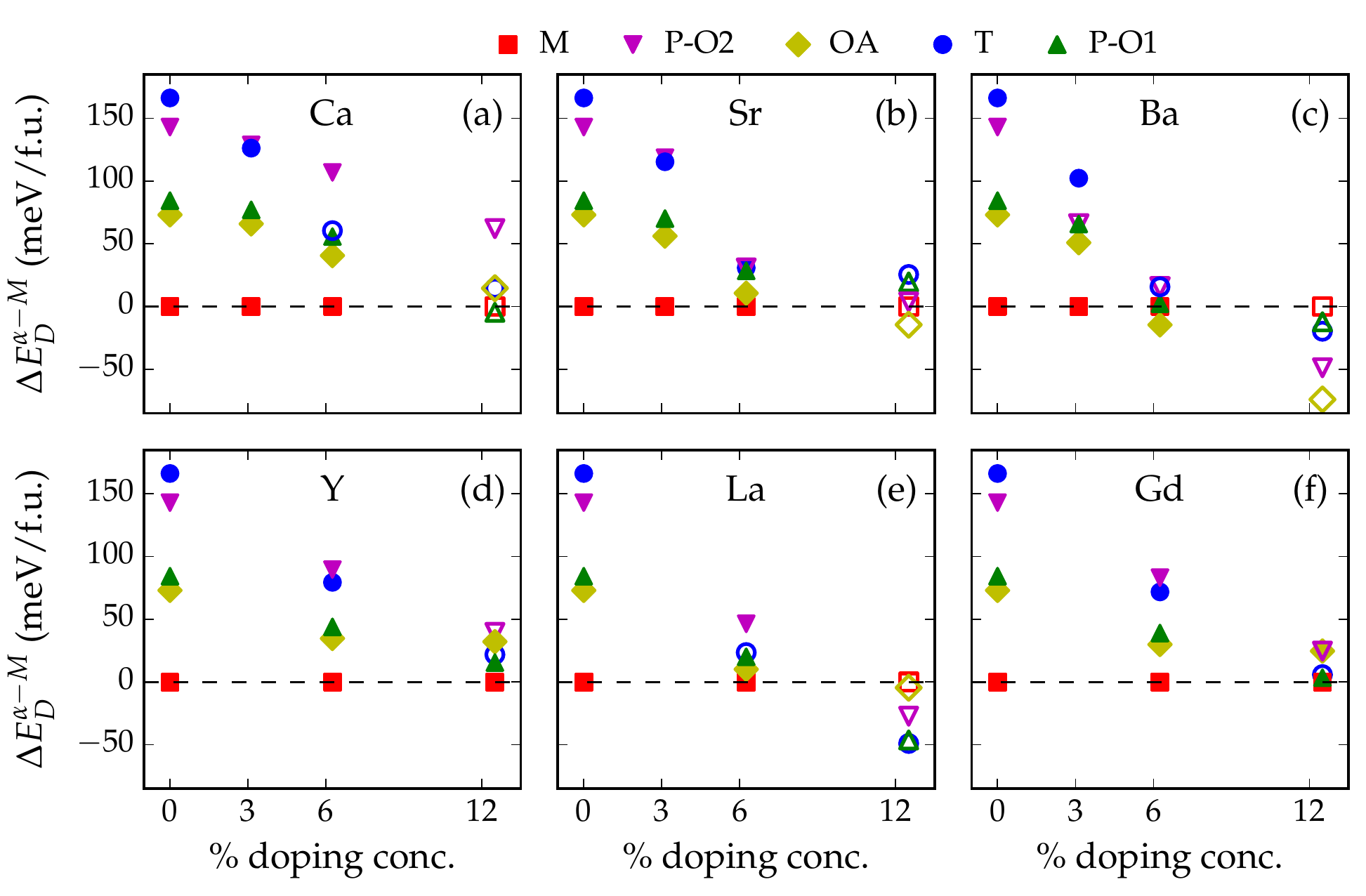}
	\caption{Phase stability of hafnia in presence of (a) Ca, (b) Sr, (c) Ba, (d) Y, (e) La, and (f) Gd, as function of their doping concentration. While the phases mostly retained their structural identity upon doping (solid symbols), in some limited cases, especially at higher doping concentration, it was hard to clearly identify the doped phases upon relaxation. Such cases are represented in open symbols based on their starting phase.}
	\label{Fig:stage3}
\end{figure}

The results of Stage 3 are presented in Fig. \ref{Fig:stage3}. We first note that while in many cases the doped hafnia phases retained their structural identity upon relaxation, there were a few cases, especially at 12.5\% doping concentration, where either it was difficult to clearly identify the doped phases or the starting phase transformed into another phase upon relaxation. We represent these unusual cases in open symbols based on their starting structure. The following key observations can be made from Fig. \ref{Fig:stage3}: (1) all of the Stage 3 dopants stabilize the P-O1 and/or the P-O2 phases with increasing doping concentration, (2) while at 3.125\% doping level, there exists substantial energy difference between the polar phases and the equilibrium M phase, at 6.25\% doping level, the P-O1 phase becomes extremely close in energy to that of the M phase, (3) at high doping concentration of 12.5\% no conclusive statements about the ground state of hafnia can be made as hafnia phases loose their structural identity at such high doping level, (4) for some doped cases, the T and even the P-O2 phase collapsed into the P-O1 phase upon relaxation, suggesting that these dopants prefer to form the relatively low energy polar P-O1 phase, and (5) between the two polar phases considered, i.e., P-O1 and P-O2, the former is clearly favored over the latter, consistent with the experimental observations of this phase.\cite{TEM_PO1_observation_hafnia}

One important limitation/assumption of the above study pertaining to the dopant and O$_{\rm vac}$ arrangement should be mentioned here. Higher doping concentration (6.25\% and 12.5\%) leads to a rather challenging modeling problem of expansion of the configurational space. For instance, for the case of 6.25\% Sr-doped hafnia, the two Sr atoms would lie on any two sites of the cation sub-lattice and the associated two O$_{\rm vac}$ on any two sites of the anion sub-lattice. Even after discounting for the symmetry of the system, a huge number of such permutations (or configurations) are possible and it is not at all trivial to determine which among them would be energetically preferred. Further, to finally determine the phase stability of doped hafnia, one would have to ascertain the lowest energy configuration of each phase. Although methods, such as, cluster expansion\cite{cluster_expansion}, etc., can be used to surmount this problem of large configurational space, these approaches are extremely computationally demanding. Nevertheless, we get some estimate of the scale of energy variations expected in our doped hafnia systems owing to the different possible configurations by computing energies of 10 diverse configurations of 6.25\% Sr-doped P-O1 phase at various dopant-dopant distances. A standard deviation of just $\sim$8 meV/f.u. in the energies of these configurations was found, suggesting that the scale of energy variations owing to different possible configurations of dopants is rather small as compared to that of the relative energies among the different phases of hafnia. Thus, we expect the trends observed in the Fig. \ref{Fig:stage3} and the conclusions made in the previous discussion to hold even when multiple possible configurations of doped hafnia phases are considered.

The results from Fig. \ref{Fig:stage3} clearly suggest that certain dopants, especially Ca, Sr, Ba, La, Y, and Gd can substantially lower the relative energy between the P-O1 and the equilibrium M phases, although no situation was encountered in which a polar phase had the lowest energy. This indicates that dopants $\it{alone}$ cannot stabilize a polar phase as the ground state in hafnia and can only $\it{assist}$ other factors, such as the surface energy, the mechanical stresses and the electric field, prevalent in the hafnia films, to form the polar phase. The disappearance of FE behavior in the absence of the aforementioned crucial factors,\cite{hafnia_review} and the empirical observation of FE behavior in pure hafnia films\cite{undoped_hafnia} further corroborates this conclusion.

\subsection{Learning from the DFT data}
\begin{figure}
	\centering
	\includegraphics[scale=0.65]{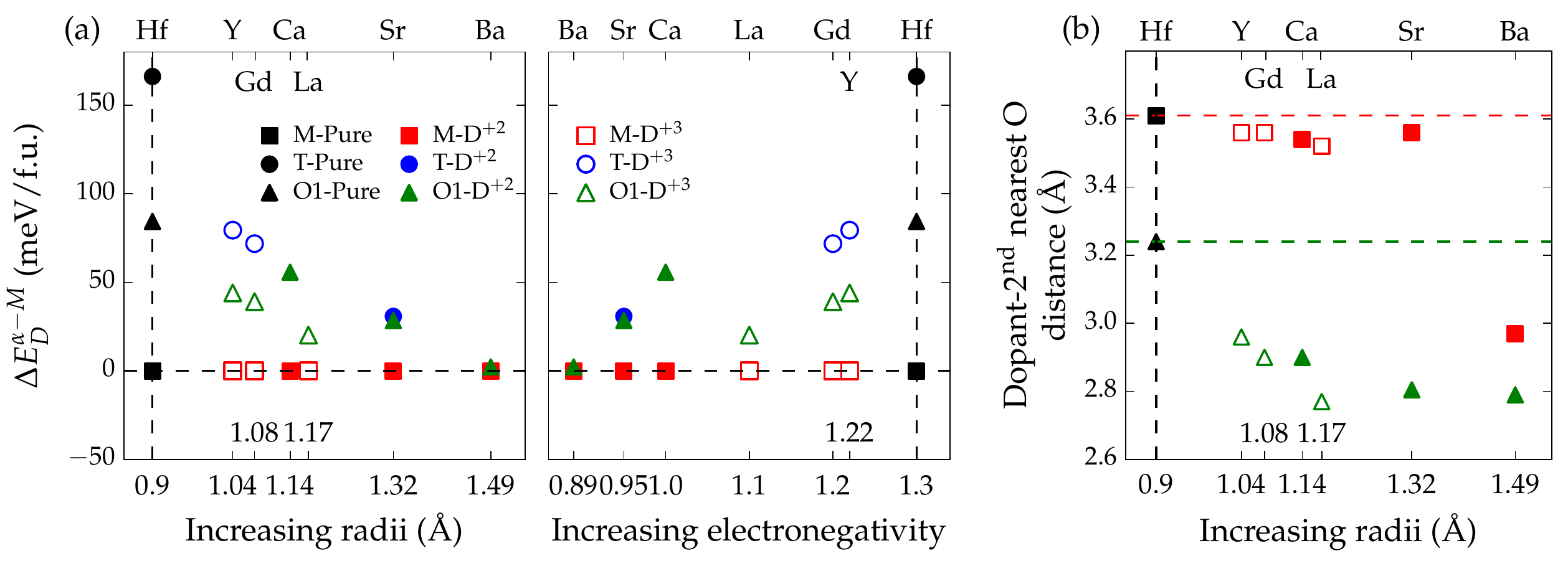}
	\caption{(a) Chemical trends in the relative energies of the M, T and P-O1 phases of hafnia with (a) ionic radius and electronegativity of a divalent (solid symbols) and trivalent (open symbols) dopant  at 6.25\% doping concentration. Some cases of the T phase collapsed into the P-O1 phase upon relaxation and are omitted here for cleanliness. (b) The distance between the dopant and the closest 2nd nearest oxygen in the case of 6.25\% doped P-O1 and M phases.}
	\label{Fig:chemical_trends}
\end{figure}

In order to reveal the dominant attributes of a dopant that help stabilize a polar phase in hafnia, we plot in Fig. \ref{Fig:chemical_trends}(a) the relative energies of the most relevant M, T and P-O1 phases against the ionic radius \cite{ionic_radii_Shannon} and the electronegativity \cite{electronegativity_pauling} of the dopants in Stage 3 for the case of 6.25\% doping level. With the dopants grouped on the basis of their valency, a clear chemical trend of dopants with \textit{higher ionic radius and lower electronegativity favoring the polar P-O1 phase} in hafnia is evident from the figure. The trend of increasing stability of the polar $Pca$2$_1$ phase with increasing dopant radii matches very well with the experimental observations \cite{dopants_hafnia_influence_RSC} of higher polarizations in hafnia systems with larger dopants. We further note that trivalent dopants considered here, owing to their ionic radii being comparable to that of Hf stabilizes the P-O1 phase at lower strains in comparison to that of the divalent dopants. Thus, trivalent dopants seem to be a superior choice to promote ferroelectricity in hafnia.

To understand the root-cause of the aforementioned chemical trends, the relaxed structures of the doped hafnia phases were carefully examined. In Fig. \ref{Fig:chemical_trends}(b), we plot the distance between the dopant and the closest 2nd nearest neighbor oxygen for the case of the M and the P-O1 phases as a function of the ionic radii of the dopants considered in Stage 3. Although this dopant-oxygen distance remains largely unaffected upon doping in the case of the M phase (with the exception of the Ba doping), it substantially reduces in the case of the P-O1 phase, suggesting formation of an additional dopant-oxygen bond. Further, as is evident from the figure, this additional bond becomes consistently shorter for dopants with larger ionic radii and lower electronegativity (not shown here). Cumulatively these observations strongly suggest formation of an energy lowering bond between the dopant cation and the 2nd nearest oxygen neighbor in the case of the P-O1 phase as the root-cause of its stabilization with respect to the M phase upon doping.

Based on the aforementioned findings and the observed chemical trends, we search the entire Periodic Table to find dopants with low electronegativity and large ionic radii that will potentially favor the polar $Pca$2$_1$ phase in hafnia. Excluding the elements studied in this work and those which are radioactive, the lanthanide series elements emerge as good dopant candidates matching these criteria. Thus, combining all the findings, results or observations from our computations we finally predict that \textit{the lanthanide series elements, the lower half of the alkaline earth metals (Ca, Sr and Ba) and Y are the most favorable dopants to promote ferroelectricity in hafnia}.

\subsection{Connection with experiments}
\begin{figure}[h]
	\centering
	\includegraphics[scale=0.6]{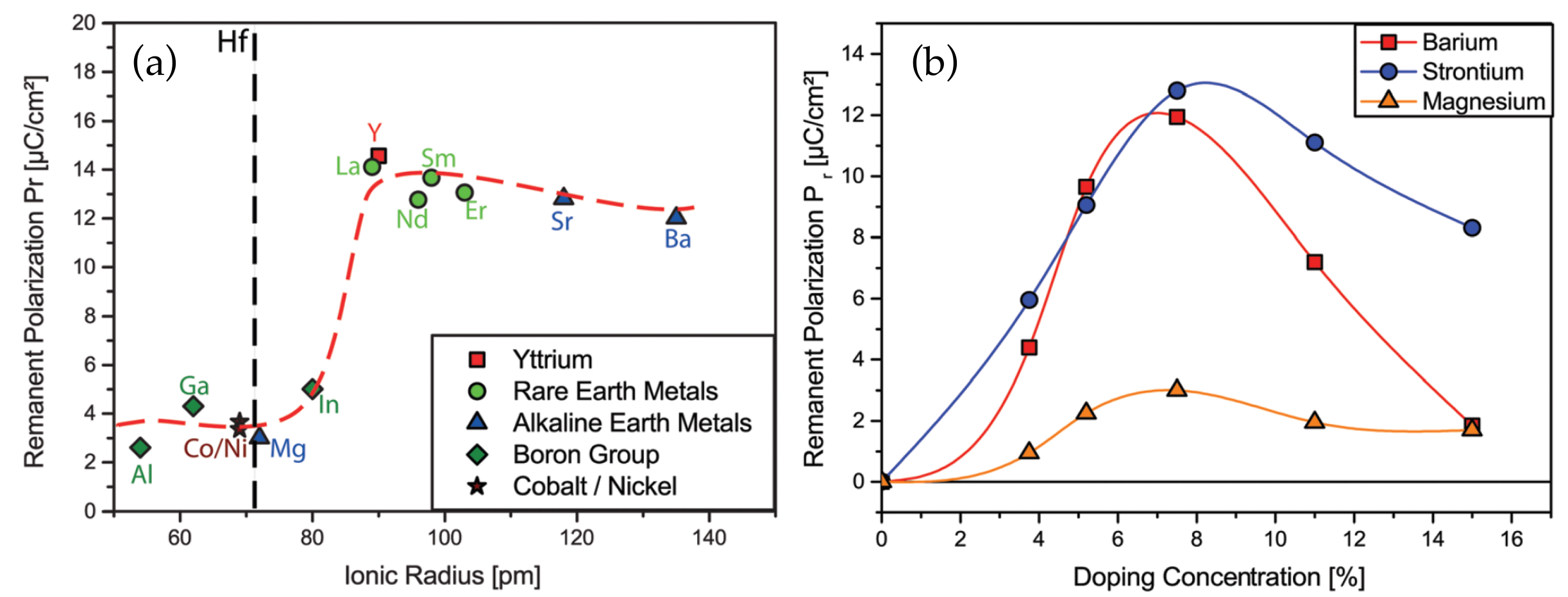}
	\caption{Trends in the measured remnant polarization of doped hafnia films with (a) dopant ionic radii and (b) doping concentration. The results are reproduced from Ref. \citenum{dopants_hafnia_influence_RSC} with permission from The Royal Society of Chemistry.}
	\label{Fig:exp_trends}
\end{figure}
Some noteworthy agreements between the theoretical predictions made in this study and the empirical observations made by Starschich et al.\cite{dopants_hafnia_influence_RSC} (major results reproduced in Fig. \ref{Fig:exp_trends}) and Schroeder et al.\cite{hafnia_dopants_effects} can also be drawn; (1) the dopants that showed substantial polarization in the empirical studies, such as Sr, Ba, Gd, Y, La were also found to stabilize the polar P-O1 phase significantly, (2) the trend of dopants of larger ionic radii stabilizing the polar P-O1 phase matches well with the experimental observation of high remnant polarization in larger dopants (see Fig. \ref{Fig:exp_trends}(a)), and (3) in agreement with the experiments, we also found that the doping concentration of 6.25\% to be most appropriate to stabilize the polar phase. As reproduced in Fig. \ref{Fig:exp_trends}(b), with increasing doping concentration, the measured polarization in hafnia films first increases, reaches a maxima around 5-8\% doping level, and then gradually decreases. Similar results are evident from this study as well. With increasing doping concentration, the polarization would initially rise due to enhanced stabilization of the polar P-O1 phase. However, after a critical doping concentration the distortions introduced in the structure would diminish the polarization of the polar phase, thus, resulting in gradual decrease in the measured polarization. Overall, the remarkable similarities between our computations and empirical observations give confidence in the assumptions made to model the hafnia systems and the predictions made in this study.

\section{Conclusions}
In summary, we investigated the influence of $\sim$40 dopants on the phase stability in hafnia using density functional theory calculations. A three stage down-selection strategy was adopted to efficiently search for promising dopants that favor the polar phases in hafnia. In Stage 1, the selected dopants were modeled under the constraints of 3.125\% substitutional doping concentration, the absence of charge neutralizing oxygen vacancy, and fixed volume. From this stage, 10 divalent and 4 trivalent dopants that favor the polar $Pca2_1$ and/or $Pmn2_1$ phase in hafnia were selected for Stage 2. While the trivalent dopants were studied directly in next stage, the divalent dopants in Stage 2 were modeled in presence of an appropriate oxygen vacancy, from which Ca, Sr and Ba were found to favor the polar $Pca2_1$ phase and were selected to Stage 3.

In Stage 3, the remaining promising candidates, i.e., Ca, Sr, Ba, Y, La and Gd doped hafnia systems were comprehensively studied at various doping concentrations with appropriate number of charge compensating oxygen vacancies. For all these dopants, increasing doping concentration enhanced the stabilization of the polar $Pca2_1$ phase. However, no case was encountered in which a polar phase becomes the ground state, suggesting that dopants $\it{alone}$ may not induce ferroelectricity in bulk hafnia and can only $\it{assist}$ other factors such as surface energy, strain, electric field, etc. Empirical measurements of relatively high remnant polarization have been made for these identified dopants, suggesting good agreement between experiments and our computations. Indeed, the doping concentration of around 5-8\% at which maximum polarization is empirically observed matches well with our predictions.

Finally, clear chemical trends of dopants with higher ionic radii and lower electronegativity favoring polar $Pca2_1$ phase in bulk hafnia were identified. For this polar phase, an additional bond between the dopant cation and the 2nd nearest oxygen neighbor was identified as the root-cause of this observation. Further, trivalent dopants, owing to their ionic radii being comparable to that of Hf, were found to favor the polar $Pca2_1$ phase at lower strains in comparison to that of the divalent dopants. Based on these insights, we were able to go beyond the dopant elements considered with the DFT calculations. We conclude that the entire lanthanide series metals, the lower half of the alkaline earth metals (Ca, Sr, Ba) and Y are the most favorable dopants to promote ferroelectricity in hafnia. These insights can be used to tailor the ferroelectric characteristics of hafnia films by selecting dopants with appropriate combination of ionic radius and electronegativity.

\begin{acknowledgement}

Financial support of this work through Grant No. W911NF-15-1-0593 from the Army Research Office (ARO) and partial computational support through a Extreme Science and Engineering Discovery Environment (XSEDE) allocation number TG-DMR080058N are acknowledged.

\end{acknowledgement}

\begin{suppinfo}
Discussion on the need of oxygen vacancy introduction in doped hafnia using electronic structure studies and the methodology adopted to characterize different phases of doped hafnia.
\end{suppinfo}

%\bibliographystyle{achemso}
%\bibliography{references}
\providecommand{\latin}[1]{#1}
\providecommand*\mcitethebibliography{\thebibliography}
\csname @ifundefined\endcsname{endmcitethebibliography}
{\let\endmcitethebibliography\endthebibliography}{}

%% graphical table of contents.
%%%%%%%%%%%%%%%%%%%%%%%%%%%%%%%%%%%%%%%%%%%%%%%%%%%%%%%%%%%%%%%%%%%%%

\newpage
\section{TOC Graphic}
\includegraphics[scale=0.57]{scheme_v6.pdf}

\end{document}